\documentclass[twocolumn,showpacs,superscriptaddress,prl]{revtex4-1}

\usepackage[dvips]{graphicx}	
\usepackage{dcolumn}			
\usepackage{bm, color, amsmath}

\begin{document}

\title{Singlet-triplet mixing in the order parameter of the noncentrosymmetric superconductor Ru$_{7}$B$_{3}$}

\author{A. S. Cameron}
\affiliation{Institut f\"ur Festk\"orper- und Materialphysik, Technische Universit\"at Dresden, D-01069 Dresden, Germany}
\author{Y. S. Yerin}
\affiliation{Institut f\"ur Festk\"orper- und Materialphysik, Technische Universit\"at Dresden, D-01069 Dresden, Germany}
\affiliation{Dipartimento di Fisica e Geologia, Universitá degli Studi di Perugia, Via Pascoli, 06123 Perugia, Italy}
\author{Y. V. Tymoshenko}
\author{P.~Y.~Portnichenko}
\affiliation{Institut f\"ur Festk\"orper- und Materialphysik, Technische Universit\"at Dresden, D-01069 Dresden, Germany}
\author{A. S. Sukhanov}
\affiliation{Institut f\"ur Festk\"orper- und Materialphysik, Technische Universit\"at Dresden, D-01069 Dresden, Germany}
\author{M. Ciomaga Hatnean}
\author{D. McK. Paul}
\author{G. Balakrishnan}
\affiliation{Department of Physics, University of Warwick, Coventry, CV47AL, United Kingdom}
\author{R. Cubitt}
\affiliation{Institut Laue-Langevin, 71 avenue des Martyrs, CS 20156, F-38042 Grenoble Cedex 9, France}
\author{A. Heinemann}
\affiliation{German Engineering Materials Science Centre (GEMS) at Heinz Maier-Leibnitz Zentrum (MLZ), Helmholtz-Zentrum Geesthacht GmbH, D-85748 Garching, Germany}
\author{D. S. Inosov}
\affiliation{Institut f\"ur Festk\"orper- und Materialphysik, Technische Universit\"at Dresden, D-01069 Dresden, Germany}

\begin{abstract}
One of the key effects which is predicted to arise in superconductors without a centre of inversion is the mixing of singlet and triplet order parameters, which are no longer good quantum numbers on their own due to parity. We have probed the gap structure in the noncentrosymmetric superconductor Ru$_7$B$_3$, through small-angle neutron diffraction from the vortex lattice, in order to search for the proposed mixed order parameter. We find that the measured temperature dependence of the vortex-lattice form factor is well characterised by a model constructed to describe the effects of broken inversion symmetry on the superconducting state, indicating the presence of a mixed singlet-triplet gap and confirming the theoretical predictions.
\end{abstract}

\date{\today}

\maketitle


Superconductivity in a noncentrosymmetric (NCS) system was first observed in CePt$_{3}$Si~\cite{Bau04}, and since the crystal structure of a NCS material breaks the spatial inversion symmetry, the nature of its superconducting order parameter can be more puzzling and complicated than in systems with a centrosymmetric crystal structure. The lack of inversion symmetry leads to parity no longer being conserved, and it has been predicted that this should lead to an unconventional pairing symmetry with the mixing of singlet and triplet pairing~\cite{Sig91, Sig07, Gor01, Bau04, Fri04,  Bau12}, and thus, to anisotropic gap structures with the possibility of accidental nodes \cite{Hay06}. Together with the antisymmetric spin-orbit coupling (ASOC), which removes the spin degeneracy of the electronic bands \cite{Rashba60}, this gives rise to the emergence of plenty of unusual and unique phenomena in the superconducting phase \cite{Karki10, Chen11, Takimoto09, Samokhin04, Yua06}. These include a possible stabilization of the Fulde-Ferrel-Larkin-Ovchinnikov state \cite{Tanaka07}, the realization of inhomogeneous peculiar helical phases \cite{Mineev94, Kau05}, the occurrence of spin currents and spontaneous magnetization at twin boundaries \cite{Arahata13}, the anomalous magnetoelectric effect with a significantly weakened paramagnetic limiting response \cite{Edelstein95, Mineev05, Fujimoto05, Mineev11, Hiasa08}, and the appearance of time-reversal symmetry breaking states \cite{Hil09, Sin14, Singh17, Biswas13, Barker15, Singh14_2, Anand11, Smidman14, Anand14, Biswas12, Bauer14, Shang18, Singh20}. However, the experimental clarification of all these effects is quite complicated, depending not only on the strength of the ASOC,  which is unambiguously present in NCS superconductors, but also on the pairing mechanism and band structure of the system. 

Among other methods, studies of vortex matter and the spatial structure of a single vortex in such compounds can be a powerful tool for the elucidation of the order parameter symmetry, the gap ``anatomy'', and other signatures and fingerprints manifested in an unconventional superconducting state with the mixing of singlet-triplet components. According to various theoretical predictions, in contrast with the vortices in conventional superconducting systems,  the lack of inversion symmetry in NCS superconductors causes exotic properties of the vortex matter. These include modulated vortex configurations without reflection symmetry, formation of bound states, clusters with non-monotonic (with multiple minima) intervortex interactions, and the creation of metastable vortex/antivortex pairs \cite{Hay06, Mat08, Hiasa09, Bau12,  Garaud20, Samoilenka20}. Therefore, vortex matter offers a unique platform to investigate not only phenomena of general importance such as domain nucleation and topology, but also the microscopic properties of superconductivity such as anisotropies of the order parameter. In other words, experimental efforts in this direction can shed light on the presence of the singlet-triplet mixing in the order parameter of NCS superconductors. 

Motivated by this challenge, and to the best our knowledge the absence of appropriate experiments, we used small-angle neutron scattering (SANS) to perform measurements of the vortex lattice (VL) of Ru$_{7}$B$_{3}$ in order to search for unconventional behavior related to broken inversion symmetry. Ru$_{7}$B$_{3}$ forms a NCS crystal structure with the space group $P6_{3}mc$~\cite{Aro59}, which is hexagonal in the basal plane. Our single-crystal sample has a superconducting transition temperature of $T_{\rm c} = 2.6$~K~\cite{Sin14}, which falls within the range of 2.5 to 3.4~K observed in earlier studies~\cite{Mat61, Fan09, Kas09}. Specific-heat and magnetisation measurements on a single crystal of Ru$_{7}$B$_{3}$ resulted in Ginzburg-Landau parameters of 21.6 and 25.5 for the $[100]$ and $[001]$ directions respectively~\cite{Kas09}, making it a reasonably strong type-II superconductor. Reports into the gap structure for Ru$_7$B$_3$ vary, with some measurements concluding an isotropic \textit{s}-wave gap~\cite{Fan09}, some a mixed gap~\cite{Dat21}, while others are less conclusive~\cite{Kas09}. Our previous investigations into the VL structure of Ru$_7$B$_3$ revealed highly unusual behaviour for magnetic fields applied along the \textbf{a} axis~\cite{Cam19}, where instead of the VL having a single or degenerate set of orientations with respect to the crystal axes for each field and temperature, which is typical behaviour seen in almost all type II superconductors, it instead showed a field-history dependent orientation. While field-history dependence of the VL structure has been seen before, for example in the two-band superconducting MgB$_{2}$ \cite{Eskildsen1, Eskildsen2} with the s-wave pairing symmetry, the type and behavior of field-history dependence observed in Ru$_7$B$_3$ appears to be unique, and we proposed a mechanism whereby it may be related to the broken time-reversal and inversion symmetries of this material. 

In this investigation, we reveal evidence of singlet-triplet mixing in the gap structure by using SANS to probe the vortex lattice. One of the key pieces of evidence for singlet-triplet mixing is the appearance of anisotropic gap structures that may contain accidental nodes~\cite{Gor01, Fri04, Hay06}, and both of these have signatures in measurements of the penetration depth as a function of temperature. SANS allows us to perform measurements of the VL form factor, which is dependent on the penetration depth and so gives us access to this quantity. In order to investigate the gap structure of Ru$_7$B$_3$ we have performed temperature-dependent measurements of the VL form factor down to 55~mK. Anisotropic gap structures and nodes have the largest effect on such observables at low temperatures, so it is important to perform measurements to low fractions of $T_{\rm c}$.


SANS measurements were performed on the D33 instrument at the Institut Laue-Langevin (ILL) in Grenoble, France~\cite{Dew08} and at the SANS-I instrument at FRM II in Garching, Germany~\cite{Hei15}. Incoming neutrons were velocity selected with a wavelength of 8~\AA, with a $\Delta \lambda / \lambda$ ratio of $\sim 10 \%$, and diffracted neutrons were detected using a position sensitive detector. The sample, a $\sim 30$ mm long cylinder with a $\sim 5$ mm diameter, was mounted on a copper holder with the \textbf{a} and \textbf{c} directions in the horizontal plane, and placed in either a dilution refrigerator (ILL) or $^{3}$He cryostat (FRM II) within a horizontal-field cryomagnet with the magnetic field applied along the \textbf{c} axis. Since $T_{\rm c}$ was above the maximum stable temperature of the dilution refrigerator, the sample was cooled in no applied field and the magnetic field was applied while at base temperature. Measurements were taken by holding the applied field and temperature constant and rocking the sample throughout all the angles that fulfil the Bragg conditions for the first-order diffraction spots of the VL. Background measurements were taken in zero field and then subtracted from the measurements in field to leave only the signal from the VL. The temperature scans were performed on warming, and diffraction patterns were treated with a Bayesian method for handling small-angle diffraction data, detailed in Ref.~\cite{Hol14}.


One of the well-established methods of determining the gap structure of a superconductor is to measure the superfluid density, $\rho_{\rm s}$, as a function of temperature. Thermal energy breaks apart Cooper pairs, forming quasiparticale excitations, and since the minimum energy to do so is determined by the superconducting gap, the form of the curve $\rho_{\rm s}(T)$ is determined by the gap structure. The superfluid density is related to the London penetration depth as $1 / \lambda^{2} \propto \rho_{\rm s}$, which is accessible through the VL form factor $F (\mathbf{q} )$, which in turn is the quantity we measure with neutrons. 

\begin{figure}
	\includegraphics[width=1\linewidth]{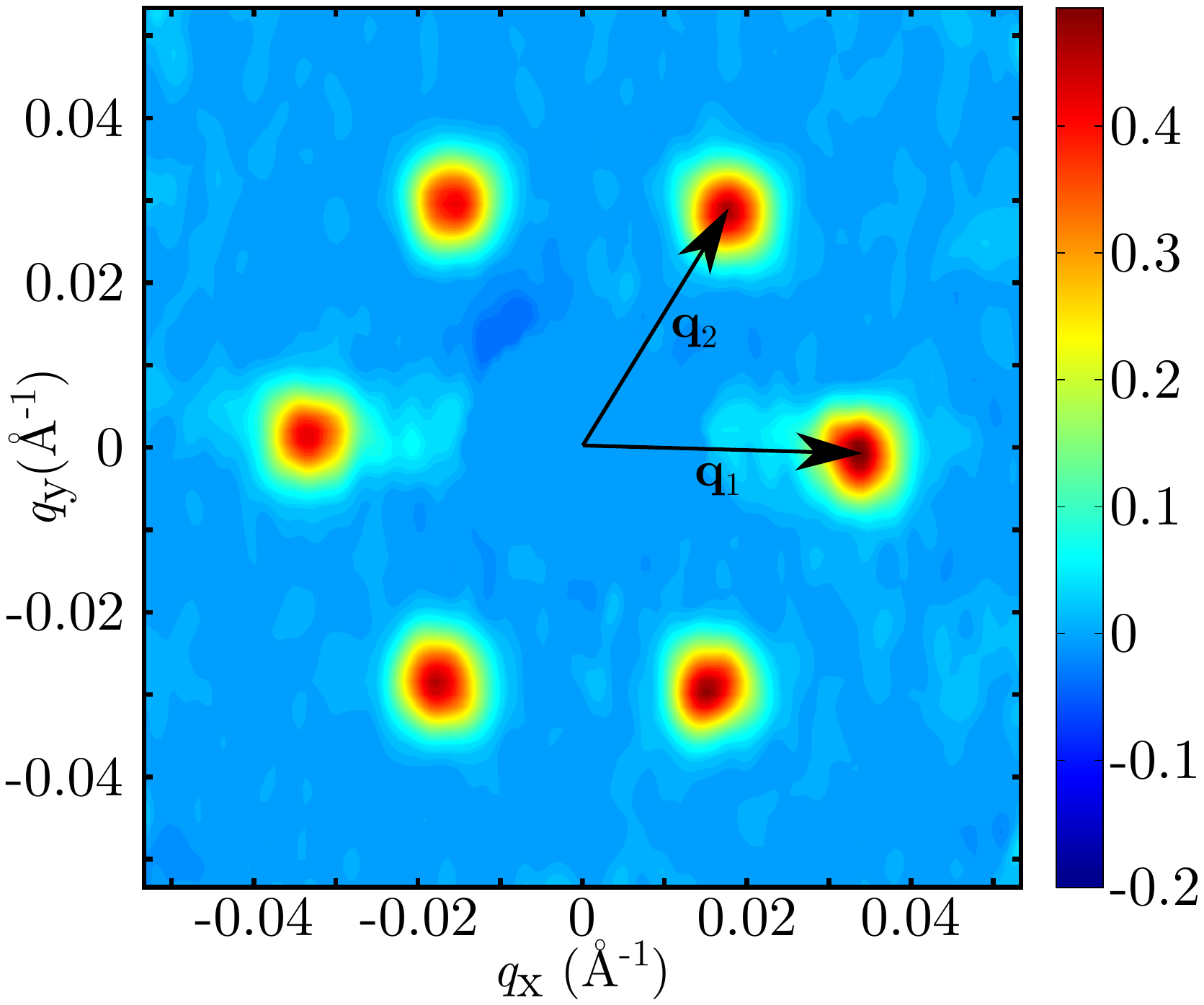}
	\caption{(Colour online) Typical diffraction pattern at 0.2~T and 55~mK for magnetic field applied along the \textbf{c}-axis. Intensity scale is in counts per measurement time, and the vectors $\mathbf{q}_{1}$ and $\mathbf{q}_2$ are the reciprocal space lattice vectors of the VL.}
		\label{Graph1}
\end{figure}

\begin{figure}
	\includegraphics[width=1\linewidth]{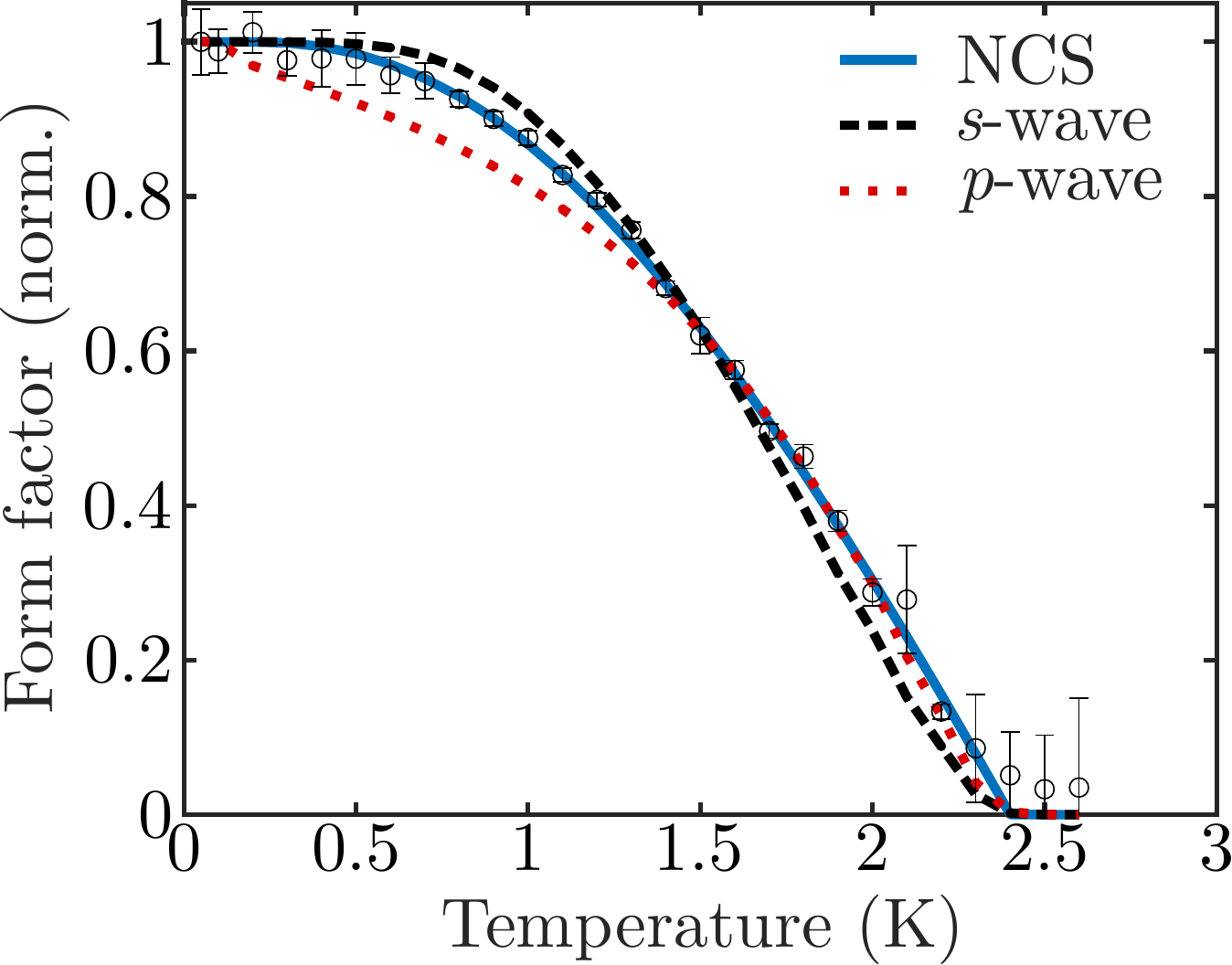}
	\caption{(Colour online) Temperature dependence of the VL form factor at 0.2~T for magnetic field applied along the \textbf{c}-axis. The solid line is a fit to the NCS model, while the dashed and dotted lines are fits to an \textit{s}-wave and \textit{p}-wave model respectively, which are described in the text.}
		\label{Graph2}
\end{figure}

The VL form factor is determined by measuring the total integrated intensity of the first order diffraction spots of the VL, and we present a typical diffraction pattern in Fig.~\ref{Graph1}. The VL form factor is related to the integrated intensity of a Bragg reflection by~\cite{Chr77}:
\begin{equation}
I_{\textbf{q}} = 2\pi V\phi \big(\frac{\gamma}{4} \big)^{2} \frac{\lambda_{n}^{2}}{\Phi_{0}^{2}\textbf{q}} |F(\textbf{q})|^{2},
\label{FF}
\end{equation}
where $V$ is the sample volume, $\phi$ is the flux of incident neutrons, $\gamma$ is the magnetic moment of the neutron in nuclear magnetons (1.91), $\lambda_{n}$ is the wavelength of the incident neutrons, $\Phi_{0}$ is the flux quantum $h/2e$, and $\mathbf{q}$ is the reciprocal space lattice vector of the VL, as indicated in Fig.~\ref{Graph1}. We present the temperature dependence of the VL form factor in Fig.~\ref{Graph2}. Data points show the experimentally measured form factor, normalised to its value at low temperature. The lines show fits to the models described in the text.

For NCS superconductors, a model of the superfluid density as a function of temperature has been developed by Hayashi~\textit{et al.}~\cite{Hay06}, from a Cooper pairing model with both spin singlet and spin triplet contributions to the order parameter. From this, they calculate the superfluid density of a NCS superconductor with the Rashba-type spin-orbit coupling. Their model describes the system in terms of the ratio of the singlet $\Psi$ to triplet $\Delta$ order parameters, $\nu = \Psi / \Delta$, the ratio of the density of states of the spin-split Fermi surfaces I and II, $\delta = (N_{\rm I} - N_{\rm II})/2N_0$, where $N_{\rm I}+N_{\rm II}=2N_0$, and the coupling constant $\lambda_{\rm m}$ which describes Cooper pair scattering between the two channels. 

Despite the fact that this model was constructed for the explanation of the power-law temperature dependence of the penetration depth in a NSC such as CePt$_3$Si, which is of the C$_4v$ point group, the approach with a two-component order parameter is valid also for Ru$_7$B$_3$ belonging to the C$_6v$ group (see e.g. \cite{Smi17} and Table 2 therein). It has been shown in Ref.~\onlinecite{Anderson84} that the lack of inversion symmetry may suppress spin-triplet pairing states because the inversion symmetry is a key ingredient for the realization of spin-triplet pairing. However, in some cases the ASOC is not devastating to the spin-triplet pairing \cite{Fri04}, and we can adopt such a model for the interpretation of experimental results in our compound. Moreover, we assume that Ru$_7$B$_3$ is a clean superconductor, and intra- and inter-component impurity effects can be neglected.

We have used the model developed by Hayashi~\textit{et al.} to fit the temperature dependence of the superfluid density in Ru$_7$B$_3$, following the method described in Ref.~\cite{Hay06}, which we present as the solid line in Fig.~\ref{Graph2}. For this fit, we have held the constant $\lambda_{\rm m} = 0.2$, as was suggested by the authors and in concordance with the usage of this model elsewhere~\cite{Yua06}. It is known that the choice of this constant does not have any qualitative effects on the results of this model~\cite{Hay06}, and we found in earlier fits that altering the value of this parameter did not significantly affect the quantitative results of the model either. The ratios of the order parameter and density of states were given a starting value of one and left free for the fit. We find that the model gives a good agreement with the data, returning values of $\nu = 2.5 \pm 0.5$ and $\delta = 0.7 \pm 0.07$. We note that the error on these numbers are reasonably large, however the qualitative effects of these parameters on the fit are not uncoupled, and so we somewhat expect a reasonably broad fit minimum in parameter space. The value of $\nu = 2.5$ corresponds to a majority spin-singlet order parameter which therefore does not possess nodes, which conforms to our expectation as nodes in the order parameter would have produced a finite slope in the form factor at $T = 0$, which we do not observe. The value of delta close to one indicates a reasonably similar, but not identical, weight of the density of states between the spin-split Fermi surfaces.

\begin{figure}
	\includegraphics[width=1\linewidth]{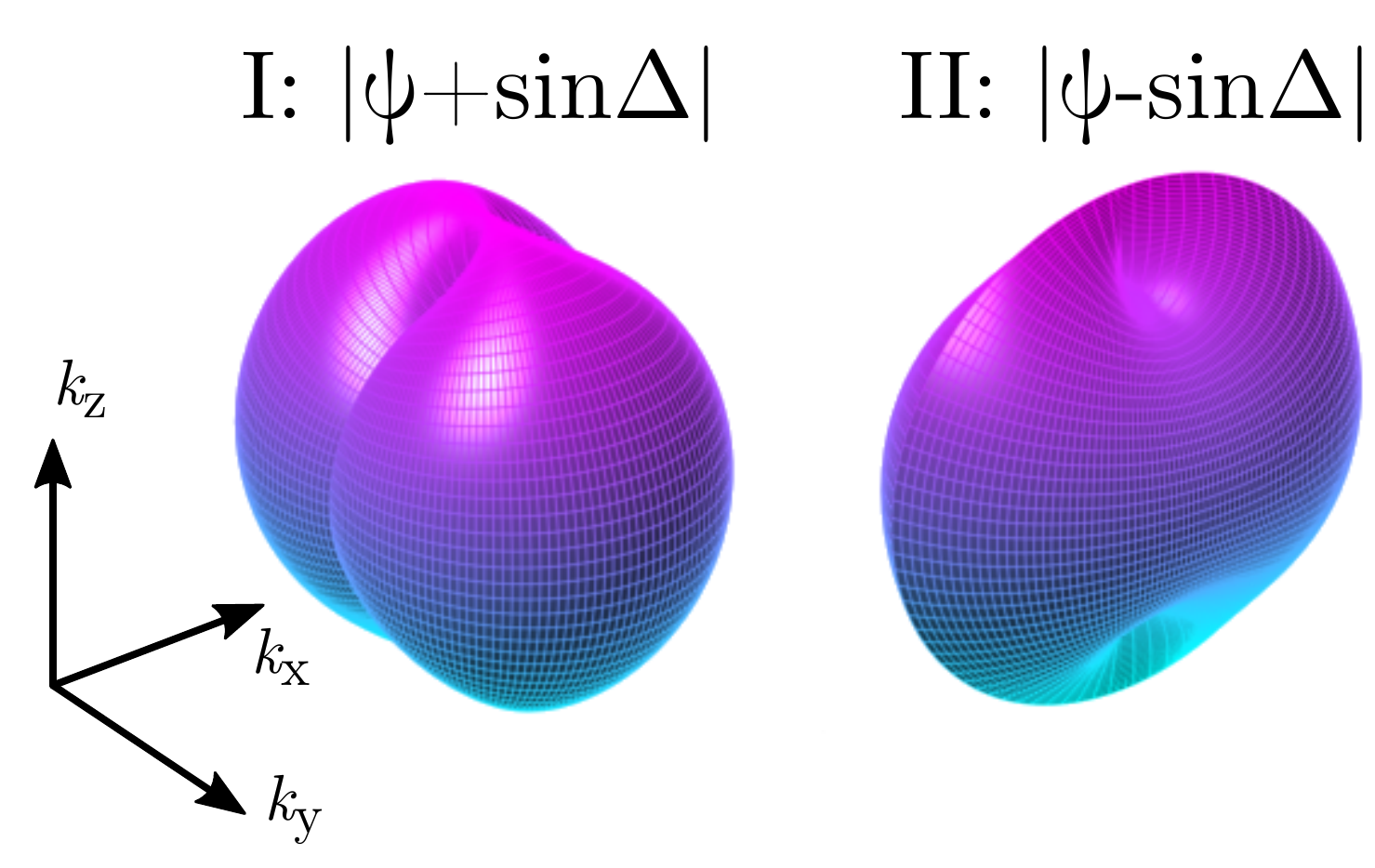}
	\caption{(Colour online) Plot of the gap functions on the two spin-spit Fermi surfaces from the fit described in the text.}
		\label{Graph3}
\end{figure}

Using the value of $\nu = 2.5$ from the fit in Fig.~\ref{Graph2}, we plot the modulus of gap values on the spin-split Fermi surfaces in Fig.~\ref{Graph3}, using the method described in Ref.~\cite{Hay06}. Here the gap on Fermi surface I is given by $\mid \psi + \Delta \sin \theta \mid$, while the gap on Fermi surface II is given by $\mid \psi - \Delta \sin \theta \mid$, where $\theta$ is the azimuthal angle in momentum space. We can see a reasonable degree of asymmetry within the gap, with gap I being extended along $k_{\rm y}$ and gap II being extended along $k_{\rm x}$. However, as expected from the value of $\nu > 1$, there is no point where the gap shrinks to zero. 


There have been a variety of models developed prior to the recent exploration of noncentrosymmetric superconductivity to describe the temperature dependence of the superfluid density for centrosymmetric systems. These models, in a similar manner to the one we use here, start with the emergence of a superconducting gap at the Fermi level, usually on a simplified approximation of the Fermi surface, and consider the effect of temperature on the gap structure and in the breaking of Cooper pairs. These models have seen a good degree of success in describing a variety of superconductors, such as the cuprates~\cite{Whi09,Whi14} and pnictides~\cite{Fur11, Ish14}. In the interest of completeness, we tested our data with these models, and examples of fits to \textit{s}-wave and \textit{p}-wave models are shown by the dashed and dotted lines, respectively, in Fig.~\ref{Graph2}. For the \textit{s}-wave model, there are two main regions of deviation between the model and the fit, one at lower temperature between 0.5 and 1.2 K, and a second between 1.7 and 2.2 K. While these deviations are not large, they are systematic. If the fit is altered so the systematic deviation from the data at low temperature is reduced, the corresponding deviation at higher temperatures worsens, and vice versa. The \textit{p}-wave model is a far worse fit than either of the other models, and while it is able to follow the data at high temperature, nodal gaps produce a characteristic finite slope at zero temperature which leads this model to be unable to fit the data at low temperature. Using the reduced chi-squared statistic for quantitative comparison, we find a value of $\chi ^2 _{\nu} = 1.4$ for the NCS model, $\chi ^2 _{\nu} = 6.5$ for the \textit{s}-wave model and $\chi ^2 _{\nu} = 11.5$ for the \textit{p}-wave model. These numbers clearly fall within the limits of accepting the NCS model and rejecting the two single-gap models. We therefore conclude that the two-component model of Hayashi~\textit{et al.}, constructed for the description of NCS systems, is appropriate for Ru$_7$B$_3$.

The returned value of $\nu = 2.5$ from our fit indicates that this is a majority singlet system, and concurs with our initial observation that the low-temperature data was inconsistent with the presence of nodes. The inability of unmixed gap models to fit the data coupled with the good agreement of the singlet-triplet mixed gap model of Hayashi~\textit{et al.} strongly suggests that singlet-triplet mixing is taking place. Measurements of the specific heat as a function of temperature and the Sommerfeld coefficient as a function of field on Ru$_7$B$_3$ by Kase~\textit{et al.} were concurrent with a ``predominance of an \textit{s}-wave channel", not outright dismissing singlet-triplet mixing but arguing for a strong dominance of the singlet component. This is therefore in agreement with the large value of $\nu$ returned by our fit and the finite gap value of Fig.~\ref{Graph3}. However, a fit to the extracted value of $H_{\rm c1}$ from magnetisation curves indicated an isotropic \textit{s}-wave gap~\cite{Fan09}. More recently, direct measurement of the gap through scanning tunneling spectroscopy measurements by Datta~\textit{et al.}~\cite{Dat21} also indicated a mixed structure appropriate for noncentrosymmetric systems. We can therefore conclude that the majority of the evidence is in favour of a mixed gap structure.


In summary, we have measured the temperature dependence of the VL form factor in Ru$_7$B$_3$ at 0.2~T. We have found that while single-component models of the form factor cannot fit our data, a model by Hayashi~\textit{et al.} constructed specifically to deal with the proposed singlet-triplet mixing predicted for NCS superconductors fits our data well, and indicates a mixed but majority singlet gap structure for the superconducting state of this material.


This project was funded by the German Research Foundation (DFG) through the research grants IN 209/3-2 and IN 209/6-1, the Research Training Group GRK 1621, and by the Federal Ministry of Education and Research (BMBF) through the projects 05K16OD2 and 05K19OD1. A.S.S. acknowledges support from the International Max Planck Research School for Chemistry and Physics of Quantum Materials (IMPRS-CPQM). Y.Y. acknowledge support by the CarESS project. The work at Warwick was supported by EPSRC, UK, through Grants EP/M028771/1 and EP/T005963/1. 

\bibliographystyle{revtest6}
\bibliography{Ru7B3}

\end{document}